\documentclass[%
 reprint,
groupedaddress,
 amsmath,amssymb,
 aps,
 physrev,
]{revtex4-2}

\usepackage{graphicx}% Include figure files
\usepackage{dcolumn}% Align table columns on decimal point
\usepackage{bm}% bold math 
\usepackage{color}
\usepackage{hyperref}% add hypertext capabilities
\hypersetup{
    colorlinks,
    citecolor=black,
    filecolor=black,
    linkcolor=black,
    urlcolor=blue
}
\usepackage{ulem}

\begin{document}

\preprint{APS/123-QED}

\title{Neutrino mass ordering in JUNO at risk from scalar NSI induced resonance}

\author{Sandhya Choubey}
\email[]{choubey@kth.se}
\author{Andreas Lund}
\email[]{alund7@kth.se}
\affiliation{Department of Physics, School of Engineering Sciences, KTH Royal Institute of Technology, AlbaNova University Center, Roslagstullsbacken 21, SE--106~91 Stockholm, Sweden}
\affiliation{The Oskar Klein Centre for Cosmoparticle Physics, AlbaNova University Center, Roslagstullsbacken 21, SE--106 91 Stockholm, Sweden}

\date{\today}

\begin{abstract}
The determination of neutrino mass ordering (NMO) is the primary goal of the currently running JUNO reactor experiment. We show that the measurement of NMO at JUNO may severely deteriorate in the presence of non-standard neutrino interactions mediated by a beyond standard model scalar (SNSI). Taking inverted ordering and the lightest neutrino mass at $m_l=0.01$ eV, the NMO sensitivity falls below $2\sigma$ for SNSI parameter values in the range $\eta_{ee}< -7.1\times 10^{-3}$ and $\eta_{ee} > 3.3\times 10^{-3}$. More importantly, for $\eta_{ee} \gtrsim 5.7\times 10^{-3}$ the NMO sensitivity in JUNO is completely lost. We show that this is due to the presence of a hitherto unrecognized resonant enhancement of the mixing angle $\theta_{12}$, which gives rise to a mass ordering degeneracy.

\end{abstract}

%\keywords{Suggested keywords}%Use showkeys class option if keyword
                              %display desired
\maketitle

\textit{Introduction} \label{sec:intro} \textemdash \, Since the firm establishment of the neutrino oscillation paradigm, the neutrino mass ordering (NMO) has been one of the most pressing ambiguities in neutrino physics. 
The intermediate baseline reactor experiment \cite{Choubey:2003qx} Jiangmen Underground Neutrino Observatory (JUNO), which started taking data in August 2025 \cite{JUNO:2025gmd}, is posed to be the first experiment to decisively determine the NMO with a projected $3\sigma$ statistical significance after 7.1 years of data-taking \cite{Abusleme_2025}. To achieve this, JUNO relies on its unprecedented precision to the reactor energy spectrum. 
This accuracy also provides the opportunity to study sub-leading effects in neutrino oscillations via various Beyond Standard Model (BSM) scenarios. Non-standard neutrino interactions mediated via scalar coupling (SNSI) is one such BSM scenario \cite{Ge_2019}. The oscillation spectrum at JUNO is expected to be sensitive to SNSI \cite{Ge_2019,Suprabh_2023}, and it is thus pertinent to ask to what extent the main design goal of JUNO, {\it{viz.}} the measurement of NMO, is impacted by SNSI. 
\begin{figure}
\begin{center}
    \includegraphics[width=0.48\textwidth]{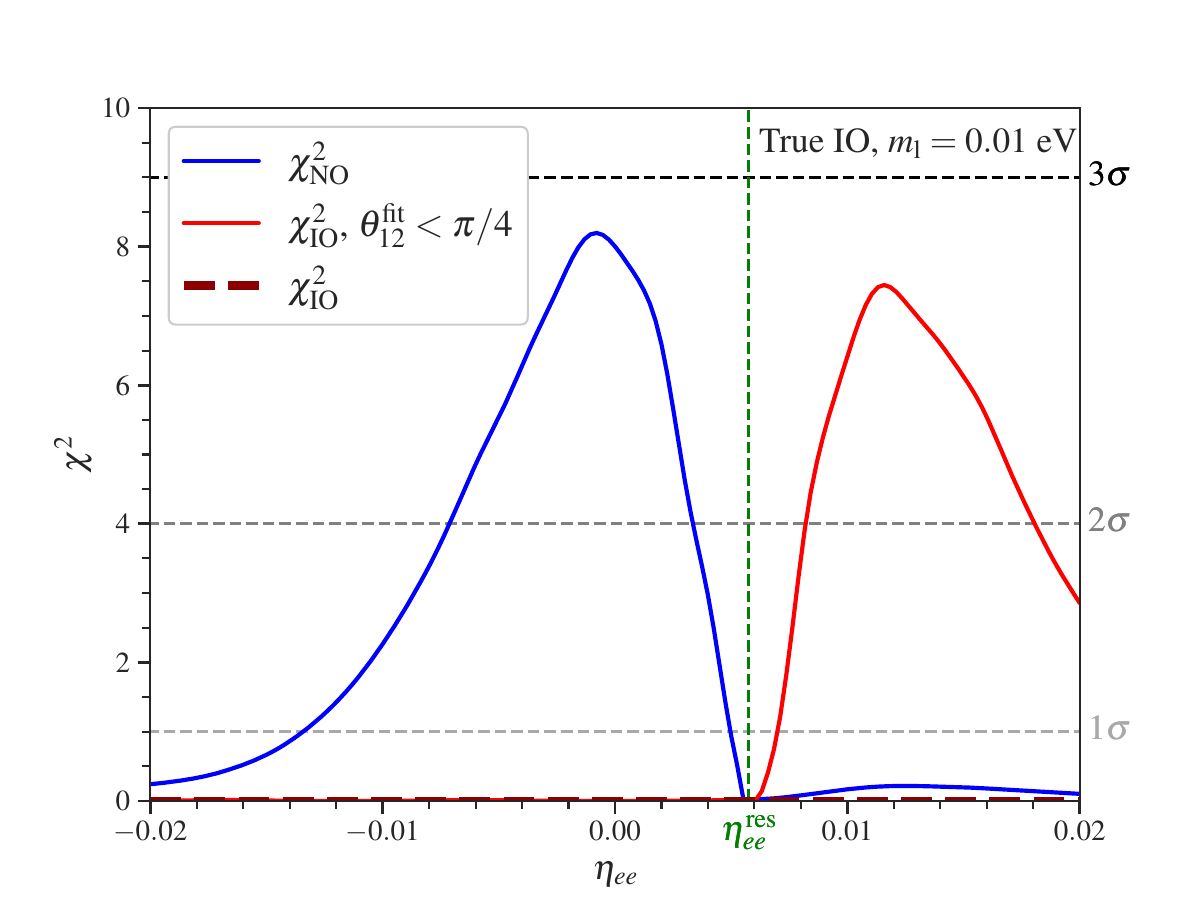}
    \caption{Neutrino mass ordering sensitivity of JUNO, when $\eta_{ee}$ is present in nature but not in the fit. Data is simulated with assumed inverted ordering (IO) and various values of the SNSI parameter $\eta_{ee}$. The lightest neutrino mass is assumed to be $m_l=0.01$ eV in all cases.}
    \label{fig:eem1vary}
\end{center}
\end{figure}

In Fig.~\ref{fig:eem1vary} we answer this question. The figure shows the expected $\chi^2$ for determining the NMO after 6.5 years of active data-taking at JUNO. 
The data is generated assuming inverted ordering and for various true values of the SNSI parameter $\eta_{ee}$, plotted along the x-axis. The simulated data is fitted separately with inverted ordering (IO) and normal ordering (NO) assuming standard oscillations ($\eta_{ee}^{fit}=0$) and the corresponding $\chi^2$ ({\it cf.~}Eq.~(\ref{eq:chi2})) are shown on the y-axis for both hypotheses. For the fit with IO, we present results for two scenarios: allowing the solar mixing angle in the fit, $\theta_{12}^{fit}$, to take any value (maroon dashed line) and allowing only $\theta_{12}^{fit}\leq \pi/4$ (red solid line). The point $\eta_{ee}=0$ in this figure corresponds to standard oscillations in nature, for which the NMO $\Delta\chi^2=\chi^2_{NO}-\chi^2_{IO}=8.0$ for ruling out the (assumed) wrong NO. This matches well with Ref.~\cite{Abusleme_2025}, where the true parameters are slightly different. 
We see from this figure that the presence of SNSI in nature results in a reduction of $\Delta \chi^2$. The NMO sensitivity falls below $2\sigma$ for $\eta_{ee}< -7.1\times 10^{-3}$ and $\eta_{ee} > 3.3\times 10^{-3}$. More importantly $\Delta \chi^2= 0$ at $\eta_{ee}=\eta_{ee}^{\rm{res}} = 5.7\times 10^{-3}$,  so the NMO sensitivity at JUNO is lost. We have $\chi^2_{NO}=0$ as well as $\chi^2_{IO}=0$ for this case. In what follows, we will show that for this value of $\eta_{ee}$ we have a resonance where the effective solar mixing angle $\theta_{12}^{\rm eff}=\pi/4$, $\grave{a}$ {\it la} MSW resonance, but for scalar NSI. 
For values of $\eta_{ee} > 5.7\times 10^{-3}$ we find that $\chi^2_{NO} \simeq 0$, while $\chi^2_{IO}$ depends on how we treat $\theta_{12}^{fit}$. If $\theta_{12}^{fit}$ is allowed to vary freely between $[0,\pi/2]$ then $\chi^2_{IO}=0$. However, if we restrict $\theta_{12}^{fit}\leq \pi/4$ and do not allow the so-called ``dark side" solutions \cite{deGouvea:2000pqg} then $\chi^2_{IO}$ is large, giving $\Delta\chi^2=\chi^2_{NO}-\chi^2_{IO} < 0$. This means that for values of $\eta_{ee} > 5.7 \times 10^{-3}$, if one does not account for the possibility of $\theta_{12}^{fit} > \pi/4$, an analysis of the JUNO data leads to favoring the wrong NMO and eventually even ruling out the correct one.

As will be shown, this loss of NMO sensitivity in JUNO in presence of SNSI is due to a resonant enhancement of $\theta_{12}$. We show that this resonance is similar to the MSW resonance of solar neutrinos; however, unlike the MSW resonance which happens due to neutrino interactions mediated by the standard model gauge boson, the resonance in our case happens due to neutrino interactions mediated by a (pseudo)scalar beyond the standard model. The resonance leads to a degeneracy in the $\bar\nu_e$ survival probabilities corresponding to NO and IO, which causes the complete loss of NMO sensitivity. \\

\textit{Simulation and statistical methods} \textemdash \, We produce Fig.~\ref{fig:eem1vary} by analyzing 6.5 years of simulated data
\cite{Abusleme_2025}. We closely follow the simulation and experimental details provided by the JUNO collaboration in \cite{Abusleme_2025, Abusleme_2022}. 
We have written a simulation code for JUNO based on a code previously developed in \cite{Du:2021rdg} that uses the GLoBES software \cite{Huber:2004ka, Huber:2007ji}. 
Our code is able to faithfully reproduce the precision and NMO sensitivity estimates provided by the JUNO collaboration in Fig.~7 of \cite{Abusleme_2022} and in Fig.~7 and Table 9 of  \cite{Abusleme_2025}.
For the statistical analysis, we generate an Asimov data set $N_{i}^{\text{true}}$ at assumed true values of the oscillation parameters taken from NuFiT6.0 \cite{Nufit6.0} and SNSI parameter $\eta_{ee}$. This data is  compared with the output of a test hypothesis $N_{i}^{\text{test}}$ using the least-squares method with a $\chi^2$ test statistic. 
Systematic uncertainties are incorporated via various pull terms \cite{Stump_2001}. The $\chi^2$ is 
\begin{align} \label{eq:chi2}
    \chi^2 =& \sum_{i} \frac{\left(N_{i}^{\text{true}}-N_{i}^{\text{test}}\right)^2}{3/(1/N_{i}^{\text{true}}+2/N_{i}^{\text{test}})} + \sum_j (\frac{x_j}{\sigma_j}),
\end{align}
where $i$ runs over 340 energy bins.
The $\chi^2$ is minimised over all oscillation parameters and systematic uncertainties. 
The rate uncertainties are composed of reactor correlated $\sigma_{\text{cor}}$ uncertainties and uncorrelated signal $\sigma_{\text{uncor}}$ uncertainties, as well as various background $\sigma_{\text{bkg}}$ uncertainties. Both signal shape uncertainties due to the near detector TAO and background shape uncertainties are incorporated as relative bin-dependent uncertainties $\sigma_{\text{sig shape}}^i$ and $\sigma_{\text{bkg shape}}^i$ \cite{Abusleme_2022}. The corresponding pull parameters $x_j=\{a,a',b,c_i,c'_i\}$ are allowed to vary freely during minimization and enter fit event rate as
\begin{equation}
    N_{i}^{\text{test}} = \sum_k (1 + a + b_k + c_i ) S_{i,k} + (1 + a' + c'_i) B_i,
\end{equation}
where $S_{i,k}$ are the signal events summed over all reactor cores $k$ and $B_i$ are the background events. We emphasize that $\eta_{ee}$ is included only in the data, $N_{i}^{\text{true}}$, generated for IO. The test hypothesis, $N_{i}^{\text{test}}$, is taken as NO with no SNSI present. Thus, our $ \chi^2$ illustrates the scenario where SNSI is present in nature but not accounted for in an analysis of the JUNO data.
Finally, the NMO sensitivity is estimated as the difference  
\begin{equation} \label{eq: NMOchi2}
    \Delta \chi^2 = \chi^2_{\text{NO}}
    -  \chi^2_{\text{IO}}.
\end{equation}
\\

\textit{Neutrino oscillations with Scalar NSI} \textemdash \, 
Neutrinos may have additional interactions in theories beyond the standard model. In particular, a BSM scalar field could generate non-standard neutrino interactions via effective terms in the Lagrangian of the form 
\begin{equation}\label{eq:Leff}
    \mathcal{L}_{snsi}
    = \frac{y_f y_{\alpha \beta}}{m_{\phi'}^2} (\bar{\nu}\nu)(\bar{f}f),
\end{equation}
where $m_{\phi'}$ is the mass of a BSM scalar field $\phi'$, $y_{\alpha\beta}$ is the Yukawa coupling of $\phi'$ with neutrinos and $y_f$ is its coupling with other fermions. As a consequence of this, neutrinos propagating in matter acquire an effective Hamiltonian, 
\begin{equation} \label{eq:Heff}
    H_{snsi}=\frac{(UMU^\dagger+\delta M)(UMU^\dagger+\delta M)^\dagger}{2 E_{\bar\nu_e}} \pm V,
\end{equation}
where $U$ is the PMNS matrix, $M=\text{diag}(m_1,m_2,m_3)$ is the neutrino mass matrix, $V$ is the standard matter potential, while $\delta M$ is the correction due to SNSI parameterized as
\begin{equation}
    \delta M = \sqrt{|\Delta m^2_{31}|} 
    \begin{pmatrix}
    \eta_{ee} & \eta_{e \mu} & \eta_{e \tau} \\
    \eta_{e \mu}^* & \eta_{\mu \mu} & \eta_{\mu \tau} \\
    \eta_{e \tau}^* & \eta_{\mu \tau}^* & \eta_{\tau \tau}
    \end{pmatrix},
\end{equation}
where $\Delta m_{ij}^2=m_i^2-m_j^2$ and 
\begin{equation}
    \eta_{\alpha \beta} = \frac{y_{\alpha \beta}}{\sqrt{|m_{31}^2|}m_{\phi'}^2} \sum_f N_f y_f,
\end{equation}
$N_f$ being the number density of fermion $f$ in matter. Note that the effective mass matrix $\tilde{M}=UMU^\dagger+\delta M$ is in general non-diagonal, even with a single SNSI parameter. Consequently, neutrino oscillations with SNSI are sensitive to the lightest neutrino mass $m_l$. For the modified Hamiltonian we can obtain the modified $\bar\nu_e$ survival probability. In this work, for illustration purposes, we only show the impact of $\eta_{ee}$ and assuming IO to be true. However, there exists similar impact of other SNSI parameters on the NMO sensitivity of JUNO and also for NO as true. Those results will be reported elsewhere. 

SNSI can be constrained from several existing experimental data. Constraints on SNSI parameters $y_{\alpha \beta}$, $y_f$ and $m_{\phi'}$ have been studied in \cite{Babu:2019iml,Sandner_2023,Venzor_2020,Babu_2019,Smirnov_2019,Dutta_2025,Das:2025zts,ESSnuSB_2023,DuttaDUNE_2024,Sarker:2023qzp,SinghaP2SO_2023,TEXONO_2025,Denton_2024}. The most stringent constraints on $\eta_{ee}$ reported so-far in the literature come from an analysis of the Borexino data \cite{Denton_2024}. There appears to be a small tension between the value of $\eta_{ee}^{\rm res}$ we show in Fig. \ref{fig:eem1vary} and the Borexino constraints. However, the constraints presented in \cite{Denton_2024} are only at the 90\% C.L. Hence, the results and subsequent conclusions drawn in this work are relevant. 

\begin{figure}[h]
\begin{center}
    \includegraphics[width=0.48\textwidth]{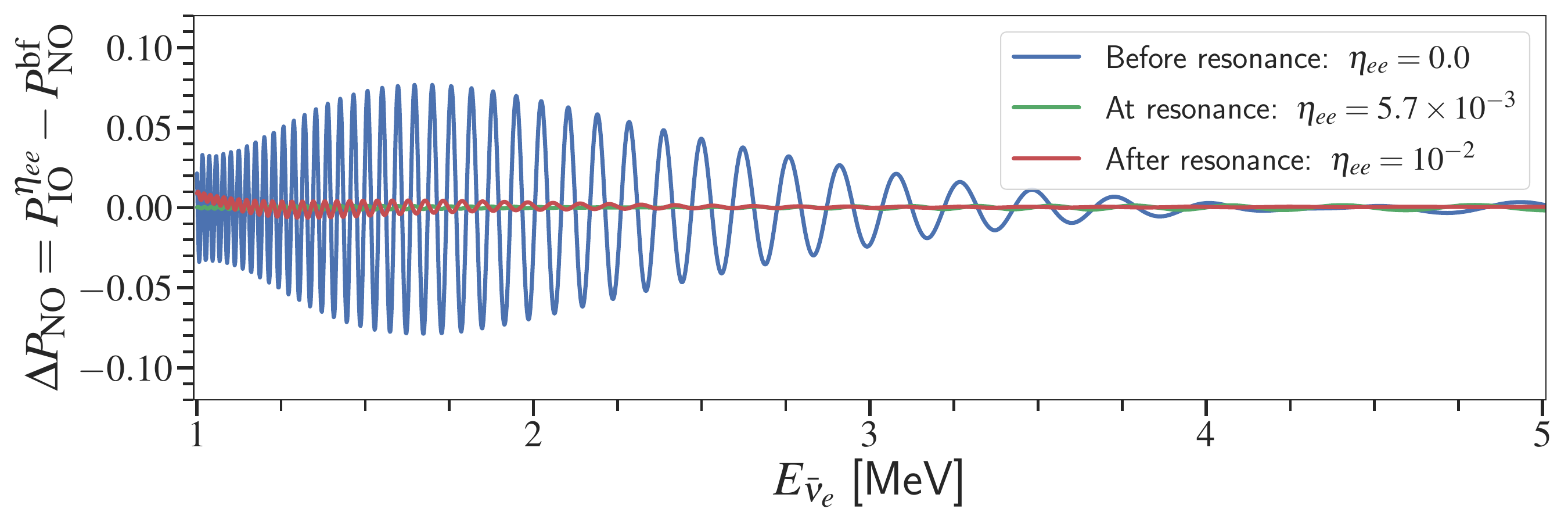}
    \includegraphics[width=0.48\textwidth]{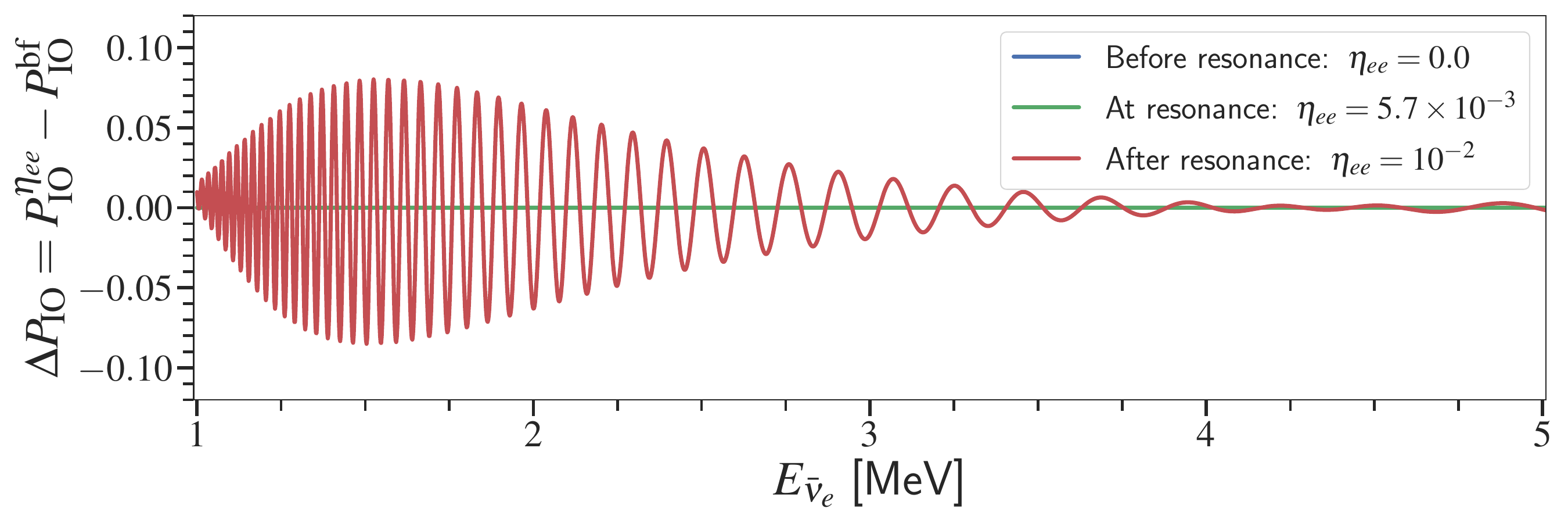}
    \caption{Probability difference between simulated data and best fit assuming NO (top-panel) and IO (bottom-panel), as function of neutrino energy. See the text for details.}
    \label{fig:degensprob}
\end{center}
\end{figure}
The loss of NMO sensitivity in JUNO in the presence of SNSI can be explained in terms of cancellations in the $\bar\nu_e$ survival probability. 
We begin by showing in Fig.~\ref{fig:degensprob} the difference in $\bar\nu_e$ survival probability in the energy range relevant for JUNO. Specifically, we show the difference in the probability between the data, which corresponds to a given value of $\eta_{ee}$ assuming IO ($P_{\rm{IO}}^{\eta_{ee}}$), and the fit where $\eta_{ee}=0$ and NMO can be NO ($P_{\rm{NO}}^{\rm{bf}}$) or IO ($P_{\rm{IO}}^{\rm{bf}}$). In each case, we show this probability difference at three values of $\eta_{ee}$: one for $\eta_{ee}=0$, one at $\eta_{ee}=\eta_{ee}^{\rm{res}}$ and one for $\eta_{ee}>\eta_{ee}^{\rm{res}}$. Recall that, in the fit, all relevant oscillation parameters are allowed to vary, and hence we compute the fit probabilities only at the values which correspond to the best-fit.

The top-panel of Fig.~\ref{fig:degensprob} shows the probability difference for the NO fit.  The $\eta_{ee}=0$ case corresponds to standard oscillations. At $\eta_{ee}^{\rm{res}}$ we see that $P_{\rm{IO}}^{\eta{ee}}-P_{\rm{NO}}^{\rm{bf}}$ goes down to exactly zero in the entire energy range of JUNO. This results in $\chi^2_{\rm{NO}}=0$ in Fig.~\ref{fig:eem1vary}. For $\eta_{ee}>\eta_{ee}^{\rm{res}}$, $P_{\rm{IO}}^{\eta_{ee}}-P_{\rm{NO}}^{\rm{bf}}$ remains small. The bottom-panel of Fig.~\ref{fig:degensprob} shows the probability difference for the IO fit with $\theta_{12}^{\rm fit}\leq \pi/4$. We do not show the IO fit case with unrestricted $\theta_{12}^{\rm fit}$, because for this case the best-fit parameters are able to align such that the probability difference is exactly zero at all energies. Here we see that while $P_{\rm{IO}}^{\eta{ee}}-P_{\rm{IO}}^{\rm{bf}}$ is zero for $\eta_{ee} \leq \eta_{ee}^{\rm{res}}$, it becomes large for $\eta_{ee}>\eta_{ee}^{\rm{res}}$, as is expected from Fig.~\ref{fig:eem1vary}. 

\begin{figure}[h]
\centering
  \includegraphics[width=.48\columnwidth]{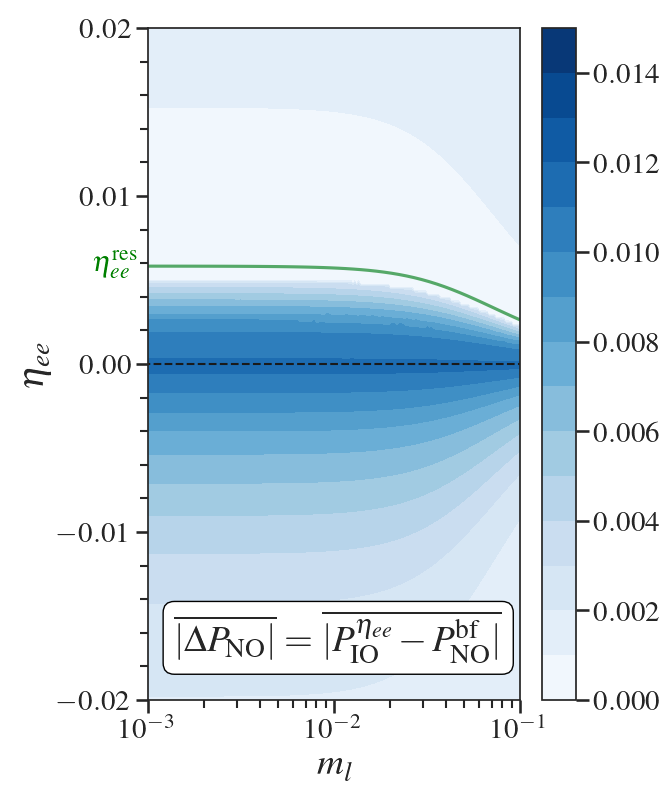}
  \includegraphics[width=.48\columnwidth]{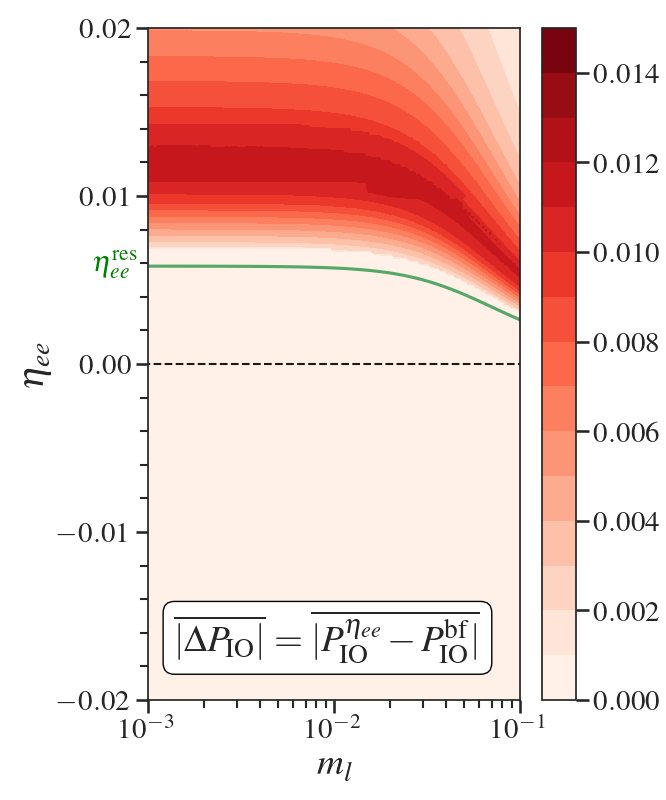}
\caption{Average integrated probability difference in the $m_l-\eta_{ee}$ plane. The right-panel shows $\overline{|\Delta P_{NO}|}=\overline{|P_{\rm{IO}}^{\eta_{ee}}-P_{\rm{NO}}^{\rm{bf}}|}$ and the left-panel shows $\overline{|\Delta P_{IO}|}=\overline{|P_{\rm{IO}}^{\eta_{ee}}-P_{\rm{IO}}^{\rm{bf}}|}$. The best fit is restricted to $\theta_{12}^{\rm{bf}}<\pi/4$.} 
\label{fig:dPetaml}
\end{figure}

We had noted earlier that the effect of SNSI on oscillations depends on $m_l$. To illustrate this, we present in the left-panel of Fig.~\ref{fig:dPetaml} the probability difference $\overline{|\Delta P_{\rm NO}|}=\overline{|P_{\rm{IO}}^{\eta_{ee}}-P_{\rm{NO}}^{\rm{bf}}|}$ averaged over energy from 1 to 9$\,$MeV. We show these as oscillograms in the $m_l-\eta_{ee}$ plane. 
As the magnitude of $\eta_{ee}$ increases from $\eta_{ee}=0$, the probability difference decreases. This implies that in presence of SNSI, JUNO starts to lose NMO sensitivity, as seen in Fig.~\ref{fig:eem1vary}. In particular, we see that for $\eta_{ee} \sim 5\times 10^{-3}$, $\overline{|\Delta P_{\rm NO}|}$ becomes very small. This is near the green line that corresponds to the theoretical value of $\eta_{ee}^{\rm{res}}$. For $\eta_{ee} > \eta_{ee}^{\rm res}$ the average probability difference remains negligible, confirming the observations from Fig.~\ref{fig:eem1vary}.
We can note the dependence of $\eta_{ee}^{\rm res}$ on $m_l$. We see that for $m_l \lesssim 10^{-2}$ eV, $\eta_{ee}^{\rm res}$ is hardly dependent on $m_l$. However, for $m_l \gtrsim 10^{-2}$ eV, $\eta_{ee}^{\rm res}$ changes somewhat with $m_l$.

The right panel of Fig.~\ref{fig:dPetaml}  shows  the energy averaged $\overline{|\Delta P_{\rm IO}|}=\overline{|P_{\rm{IO}}^{\eta_{ee}}-P_{\rm{IO}}^{\rm{bf}}|}$, where $\theta_{12}^{\rm fit}\leq \pi/4$. Here we see that for $\eta_{ee} \leq \eta_{ee}^{\rm res}$, the probability difference is essentially zero. However, for $\eta_{ee} > \eta_{ee}^{\rm res}$, the probability difference increases from zero, reaching a maximum around $2\eta_{ee}^{\rm res}$, thereafter falling again. This is completely in line with what was seen in Fig.~\ref{fig:eem1vary}.
\\

\begin{figure}[]
\begin{center}
    \includegraphics[width=0.48\textwidth]{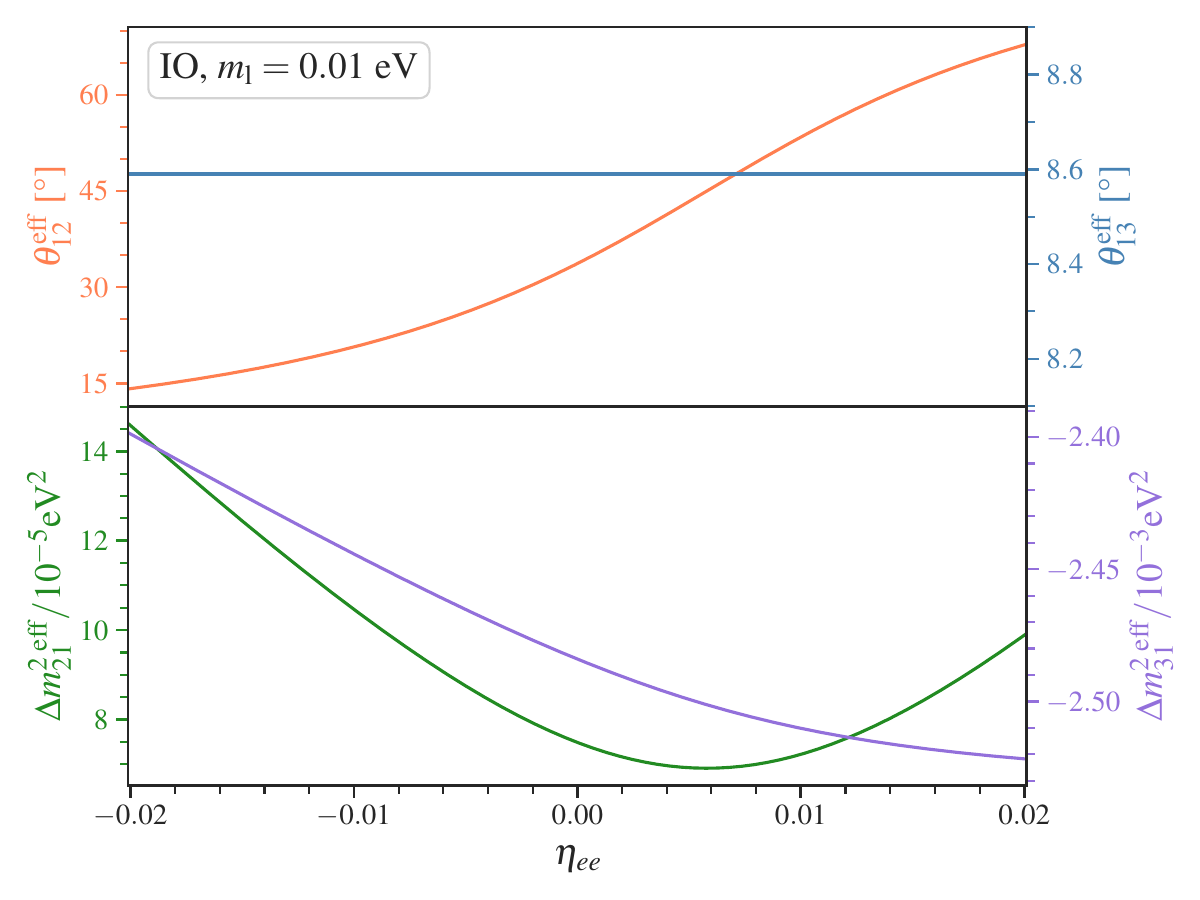}
    \caption{Effective oscillation parameters as functions of $\eta_{ee}$.}
    \label{fig:paramsfitvstrue}
\end{center}
\end{figure}

\textit{The $\eta_{ee}$ induced resonance} \label{sec:analytic} \textemdash \,
In order to understand the resonance due to SNSI that leads to the degeneracy in the probabilities, we present a simplified analytical formulation. We derive these formulae by employing a change of basis approach with the Jacobi method \cite{Ioannisian_2018,Suprabh_2023}. 
We initially rotate $H_{snsi}$ as $H'=U_a^\dagger H_{snsi} U_a$, where $U_a=R_{23} R_\delta R_{13}$. This rotation also appears in the amplitude as $S=e^{-i H_{snsi} L}=U_a e^{-i H' L} U_a^\dagger$. We found that for small $\eta_{ee}$, only one SNSI term in $H'$ contributes significantly to the probability. Therefore, we simplify our analytical calculations by taking
\begin{align} \label{eq:HamiltonianRotated}
    &H_{11}' \!=\! m_1^2 c_{12}^2 + m_2^2 s_{12}^2 + 2 \eta_{ee} \sqrt{|\Delta m_{31}^2|}  c_{13}^2(m_1 c_{12}^2 + m_2 s_{12}^2), \nonumber \\ &
    H_{12}' = s_{12} c_{12} \Delta m_{21}^2 , ~
    H_{13}' = 0, \nonumber \\ &
    H_{22}' = m_1^2 s_{12}^2 + m_2^2 c_{12}^2 ,~
    H_{23}' = 0, ~
    H_{33}' = m_3^2,
\end{align}
where $c_{ij}=\cos\theta_{ij}$ and $s_{ij}=\sin\theta_{ij}$. 
By diagonalizing this matrix with two consecutive Jacobi rotations, one acquires simple expressions for the effective oscillation parameters in the presence of SNSI:
\begin{align}
    & \tan2\theta_{12}^{\text{eff}}= \frac{\Delta m_{21}^2 \sin2\theta_{12}} { \Delta m_{21}^2 \cos2\theta_{12} - \eta_{ee} B},\label{eq:t12eff} \\
    & {\Delta m^2_{21}}\!\!^{\text{eff}} = \Delta m_{21}^2\cos2(\theta_{12}^{\text{eff}}-\theta_{12})  \!-\! \eta_{ee} B\cos2\theta_{12}^{\text{eff}}, \label{eq:dm21eff}\\
    & {\Delta m^2_{31}}\!\!^{\text{eff}} = \Delta m_{31}^2 \!-\! \Delta m_{21}^2 \sin^2(\theta_{12}^{\text{eff}}\!-\!\theta_{12})  \!-\! \eta_{ee} B{c_{12}^{\text{eff}}}^2 ,\label{eq:dm31eff}
\end{align}
where $B=2\sqrt{|\Delta m_{31}^2|} c_{13}^2 (m_1 c_{12}^2  + m_2 s_{12}^2)$. Note that our simplified formulae yield $\theta_{13}^{\rm{eff}}=\theta_{13}$, whereas in the full picture one would find that $\theta_{13}^{\rm{eff}}$ varies slightly. We show in Fig.~\ref{fig:paramsfitvstrue} these effective oscillation parameters as a function of $\eta_{ee}$. The top panel shows the mixing angles $\theta_{12}^{\text{eff}}$ and $\theta_{13}^{\text{eff}}$, while the bottom panel shows the mass squared differences ${\Delta m^2_{21}}\!\!^{\text{eff}}$ and ${\Delta m^2_{31}}\!\!^{\text{eff}}$. We see that ${|\Delta m^2_{31}}\!\!^{\text{eff}}|$ increases with $\eta_{ee}$, while ${\Delta m^2_{21}}\!\!^{\text{eff}}$ decreases and then increases. And finally, we see that $\theta_{12}^{\text{eff}}$ increases monotonically with $\eta_{ee}$, reaching the value $\pi/4$, due to the SNSI-resonance. This resonant behavior can be seen in Eq.~(\ref{eq:t12eff}), where the denominator vanishes when
\begin{align}
\Delta m_{21}^2 \cos2\theta_{12} = \eta_{ee}^{\rm res} B,
\label{eq:res}
\end{align}
resulting in $\theta_{12}^{\text{eff}}=\pi/4$. {\it This is the SNSI-resonance condition.} Note that this resonance is similar to the MSW resonance, where the term $\eta_{ee} B$ plays the same role as the standard matter term $A=2VE_{\bar\nu_e}$ does in MSW resonance. At this SNSI-resonance, ${\Delta m^2_{21}}\!\!^{\text{eff}} = \Delta m_{21}^2 \sin2\theta_{12}$, which is its minimum value in presence of SNSI. Further note that for $\eta_{ee}>\eta_{ee}^{\rm res}$, $\theta_{12}^{\text{eff}}$ continues to increase, and will asymptotically reach its maximum value $\pi/2$ for very large values of $\eta_{ee}$ (not presented in this figure). The range $[\pi/4-\pi/2]$ constitutes the ``dark side" solution for the solar mixing angle. Interestingly, at $\eta_{ee}=2 \eta_{ee}^{\rm{res}}$ one finds $\theta_{12}^{\text{eff}}=\pi/2-\theta_{12}$ and ${\Delta m^2_{21}}\!\!^{\text{eff}} = \Delta m_{21}^2$. 

The $\bar\nu_e$ survival probability is given as
\begin{align} \label{eq:Pee}
    P =& 1-\sin^2 2 \theta_{12}c_{13}^4 \sin^2 \Delta_{21} 
    \nonumber
    \\ &-\sin^2 2 \theta_{13} \bigg [ c_{12}^2 \sin^2 \Delta_{31} + s_{12}^2 \sin^2\Delta_{32} \bigg ],
\end{align}
where $\Delta_{ij}=\Delta m^2_{ij}L/4E$. $P_{\rm{NO}}^{\rm{bf}}$ can be computed by inserting in Eq.~(\ref{eq:Pee}) the best-fit parameters coming from the NO fit and $P_{\rm{IO}}^{\eta_{ee}}$ can be computed by using IO and replacing the oscillation parameters in Eq.~(\ref{eq:Pee}) with their corresponding effective parameters given in Eqs.~(\ref{eq:t12eff}-\ref{eq:dm31eff}). Note that the second term in Eq.~(\ref{eq:Pee}) that depends on $\Delta m^2_{31}$ and its sign is the sub-leading term, while the $\Delta m^2_{21}$-driven first  term is the leading term. Hence, the fit always chooses $\Delta_{21}^{\rm bf} \approx \Delta_{21}^{\text{eff}}$ and $\sin^2 2 \theta_{12}^{\rm bf} \approx \sin^2 2 \theta_{12}^{\rm{eff}}$. The latter allows two best-fit values for the solar mixing angle, $\theta_{12}^{\rm bf}=\theta_{12}^{\rm eff}$ and $\theta_{12}^{\rm bf} = \pi/2-\theta_{12}^{\rm eff}$. Note from Fig.~\ref{fig:paramsfitvstrue} at $\eta_{ee}^{\rm res}$, $\theta_{12}^{\rm{eff}}=\pi/4$ and beyond that $\theta_{12}^{\rm{eff}}>\pi/4$. Note also that the NMO dependent last term of Eq.~(\ref{eq:Pee}) depends on the octant of $\theta_{12}$. Therefore for $\eta_{ee} \geq \eta_{ee}^{\rm res}$, when we fit the data for IO with the NO hypothesis, the fit chooses $\theta_{12}^{\rm bf} = \pi/2-\theta_{12}^{\rm eff}$ and the probability difference becomes 
\begin{align} \label{eq:dPsimple}
    \Delta P^{\rm NO} &=   P_{\rm{IO}}^{\eta_{ee}}-P_{\rm{NO}}^{\rm{bf}} = \nonumber
    \\ &-\sin^22 \theta_{13} \bigg[ {c_{12}^{\rm {eff}}}^2 \left(\sin^2\Delta_{31}^{\rm {eff}} - \sin^2\Delta_{32}^{\rm bf} \right) +  \nonumber
    \\ & \quad \quad \quad \quad \quad \quad  {s_{12}^{\rm {eff}}}^2 \left(\sin^2\Delta_{32}^{\rm {eff}} - \sin^2\Delta_{31}^{\rm  bf}\right) \bigg ].
\end{align}
Finally, for $\Delta P=0$ one needs $\sin^2\Delta_{32}^{bf}=\sin^2\Delta_{31}^{\rm {eff}}$ and $\sin^2\Delta_{31}^{bf}=\sin^2\Delta_{32}^{\rm {eff}}$ to be simultaneously satisfied in the fit. That is indeed  the case, as can be seen from Eqs.~(\ref{eq:t12eff}-\ref{eq:dm31eff}) and holds for all values of $\eta_{ee} \geq \eta_{ee}^{\rm res}$. Note that while the argument above gives $\Delta P^{\rm NO}=0$ exactly, we have seen in Figs.~\ref{fig:eem1vary} and Fig.~\ref{fig:degensprob} that the degeneracy is not exact for $\eta_{ee} > \eta_{ee}^{\rm res}$. This is because for the analytical formulation we have used the approximation given in Eq.~(\ref{eq:HamiltonianRotated}), while the earlier results were obtained using the exact numerical computations using the full Hamiltonian. 

The IO fit case can be explained as follows. Since the mass ordering is the same in data and in theory, it is always possible for the fit to match the data by tuning the oscillation parameters in the fit to the values corresponding to the effective parameters ({\it cf.}~Eqs.(\ref{eq:t12eff}-\ref{eq:dm31eff})). This is also required  for the solar mixing angle, for which $\theta_{12}^{\rm eff}$ goes to the dark side for $\eta_{ee} \geq \eta_{ee}^{\rm res}$. Hence, for the IO fit $\chi^2_{\rm IO}\approx 0$ for the $\theta_{12}^{\rm fit}$ completely free case, while for the $\theta_{12}^{\rm fit} >\pi/4 $ case, $\chi^2_{\rm IO} \neq 0$ and large.  \\

\textit{Conclusion} \textemdash \, 
In this work, we considered the impact of scalar-mediated non-standard neutrino interactions (SNSI) on neutrino oscillations. We showed that the mass ordering sensitivity of JUNO is severely compromised in the presence of SNSI. In particular, we find that the NMO sensitivity falls below $2\sigma$ for the SNSI parameter values in the range $\eta_{ee}< -7.1\times 10^{-3}$ and $\eta_{ee} > 3.3\times 10^{-3}$. We also showed that SNSI bring a resonance in the solar mixing sector for a certain $\eta_{ee}^{\rm res}$. There is an exact degeneracy between the probability with IO at $\eta_{ee} = \eta_{ee}^{\rm res}$, and the probability with NO at $\eta_{ee}=0$. For $\eta_{ee} > \eta_{ee}^{\rm res}$ this degeneracy still holds approximately. Hence, for any $\eta_{ee} \geq \eta_{ee}^{\rm res}$ the NMO sensitivity of JUNO vanishes.

While we only showed results for the SNSI parameter $\eta_{ee}$, a comparable loss of NMO sensitivity was seen for other SNSI parameters. Furthermore, the NMO sensitivity similarly vanishes if one takes NO in the data and IO in the fit. These results are qualitatively analogous to those presented here and will be reported elsewhere. 

%\onecolumn
%\begin{acknowledgments}
\onecolumngrid
\textit{Acknowledgement} \textemdash \, 
We extend our sincere gratitude to Sampsa Vihonen for collaboration in the early stages of this work. This work is supported by the Swedish Research Council (Vetenskapsrådet) through grant 2023-05141. 
%\end{acknowledgments}
%\twocolumn

\bibliography{references}% Produces the bibliography via BibTeX.

\end{document}